\documentclass[aps,prb,showpacs,citeautoscript,superscriptaddress,twocolumn]{revtex4}

\usepackage{graphicx}
\usepackage{amsmath}
\usepackage{amssymb}
\usepackage{color}

\newcommand{\bea}{\begin{eqnarray*}}
\newcommand{\eea}{\end{eqnarray*}}
\newcommand{\bne}{\begin{equation*}}
\newcommand{\ede}{\end{equation*}}

\newcommand{\bnen}{\begin{equation}}
\newcommand{\eden}{\end{equation}}
\newcommand{\bean}{\begin{eqnarray}}
\newcommand{\eean}{\end{eqnarray}}
\newcommand{\bnsn}{\begin{subequations}}
\newcommand{\edsn}{\end{subequations}}

\newcommand{\bna}{\begin{array}}
\newcommand{\eda}{\end{array}}
\newcommand{\bnm}{\begin{enumerate}}
\newcommand{\edm}{\end{enumerate}}
\newcommand{\bni}{\begin{itemize}}
\newcommand{\edi}{\end{itemize}}

\newcommand{\avg}[1]{\langle #1 \rangle}

\newcommand{\ket}[1]{| #1 \rangle}
\newcommand{\bra}[1]{\langle #1 |}

\begin{document}

\title{Maximal Rabi frequency of an electrically driven spin in a disordered
magnetic field}

\author{G\'abor Sz\'echenyi}
\affiliation{Institute of Physics, E\"otv\"os University, Budapest, Hungary}

\author{Andr\'as P\'alyi}
\affiliation{Institute of Physics, E\"otv\"os University, Budapest, Hungary}
\affiliation{MTA-BME Exotic Quantum Phases Research Group,
Budapest University of Technology and Economics, Budapest, Hungary}

\date{\today}

\newcommand{\Dso}{\Delta_{\rm SO}}
\begin{abstract}
We present a theoretical study of the spin dynamics of a single electron
confined in a quantum dot.
Spin dynamics is induced by the interplay of electrical driving and
the presence of a spatially disordered magnetic field,
the latter being transverse to a homogeneous magnetic field.
We focus on the case of strong driving, i.e., when the oscillation
amplitude $A$ of the electron's wave packet is comparable to the
quantum dot length $L$.
We show that electrically driven spin resonance can be induced
in this system by subharmonic driving, i.e., if the excitation
frequency is an integer fraction (1/2, 1/3, etc) of the Larmor
frequency.
At strong driving we find that
(i) the 
Rabi frequencies at the subharmonic resonances 
are comparable to the Rabi frequency at the fundamental resonance, and
(ii) at each subharmonic resonance, the Rabi frequency can be 
maximized 
by setting the drive strength to an optimal, finite value.
In the context of practical quantum information processing, these findings 
highlight the availability of subharmonic resonances for qubit control with 
effectivity close to that of the fundamental resonance, and
the possibility that increasing the drive strength might lead to
a decreasing qubit-flip speed.
Our simple model is applied to describe electrical control of 
a spin-valley qubit in a weakly disordered carbon nanotube.
\end{abstract}

\maketitle


\emph{Introduction.}
Controlled two-level systems are essential constituents of a number
of existing applications including magnetic resonance imaging and atomic 
clocks, and could form the basis for the potential future technology
of quantum information processing.
An archetype of controlled two-level systems is the 
electron spin in the presence of a static magnetic field,
illuminated by an ac transverse magnetic field\cite{Shirley,Koppens-esr}.
If the energy quantum $\hbar \omega$ 
of the ac field matches the energy distance $\hbar \omega_L$ between
the two spin levels, then the electron, occupying
the ground state before switching on the radiation, will evolve coherently
and cyclically between the ground and excited spin states.
This dynamics is known as Rabi oscillation, and the inverse time scale 
of a complete ground state---excited state transition, usually
proportional to the amplitude $B_{\rm ac}$
of the ac field,  is called the Rabi frequency.

Rabi oscillations can also occur if the energy quantum of the 
transverse ac field is an integer fraction (\emph{subharmonic}) of 
the energy difference between the spin levels\cite{Shirley}, i.e., 
if $\hbar \omega = \hbar \omega_L/N$ with
$N \in \mathbb{Z}^+$.
As long as the spin splitting $\hbar \omega_L$ dominates the
amplitude $B_{\rm ac}$ of the transverse field, 
the Rabi frequency $\Omega^{(N)}$ of the 
$N$th subharmonic transition 
(a.k.a. the $N$-photon transition)
shows the dependence
$\Omega^{(N)}\propto B_{\rm ac} \left(\frac{B_{\rm ac}}{\hbar \omega_L}\right)^{(N-1)}$,
meaning that
(i) a higher $N$ implies a slower Rabi oscillation, and 
(ii) the Rabi frequency increases when $B_{\rm ac}$ is increased.

Subharmonic spin resonances have recently been 
observed\cite{Laird-sst,Schroer,NadjPerge,FeiPei,Laird} in 
\emph{electrically}
driven quantum dots 
(QDs)\cite{Nowack-esr,Laird-prl,PioroLadriere-esr,Golovach,Flindt}.
In these systems, electrical driving might be 
favorable over magnetic excitation,
as the former allows for selective addressing of spin-based quantum bits in a
multi-quantum-dot register.
The ac electric field induces an oscillatory motion of the electron in the QD,
which in turn gives rise to spin rotations, provided that a 
suitable interaction between spin and motion is present in the sample. 
Examples are
spin-orbit and hyperfine interactions, and spatially dependent 
magnetic fields.
Theory works have addressed subharmonic 
EDSR (electrically driven spin resonance) assisted by $g$-tensor modulation for 
donor-bound electrons\cite{De} and holes in self-assembled QDs 
\cite{Pingenot}, and by
nonlinear charge dynamics in a double-dot 
potential\cite{Rashba-subharmonic-edsr}.
References \onlinecite{Nowak,Osika} analyzed subharmonic transitions via 
numerical simulations of EDSR in nanowire QDs.

\begin{figure}
\includegraphics[width=0.4\textwidth]{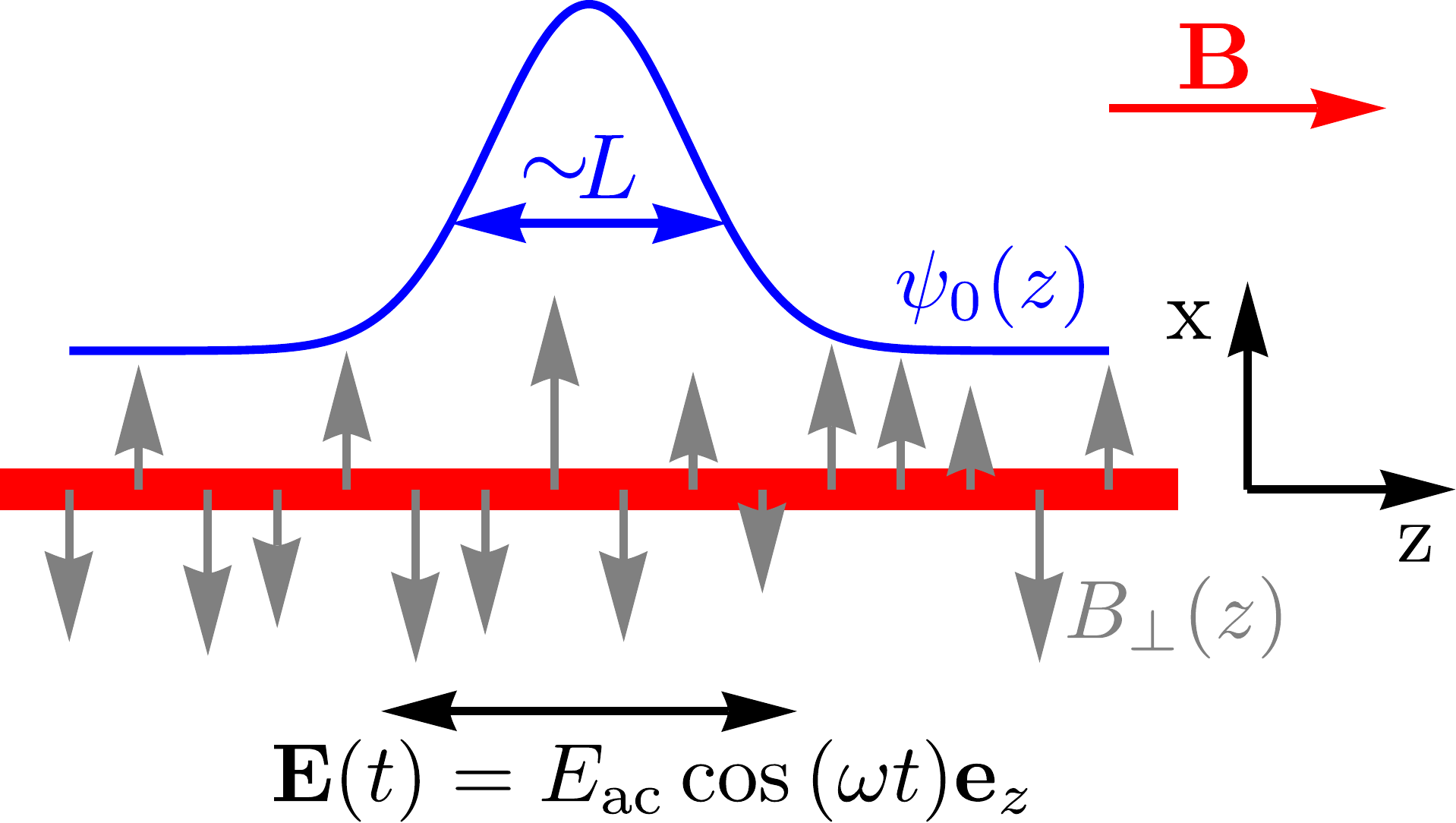}
\caption{\label{fig:setup} 
(Color online)
A simple model for electrically driven spin resonance. 
An electron confined to a quantum dot in a nanowire,
subject to a static homogeneous $B$-field and a spatially inhomogeneous, disordered transverse magnetic field (represented by the gray
vertical arrows)
is shaken by the ac $E$-field $\mathbf E(t)$.
}\end{figure}

In this work, we consider a practically relevant, yet simple model of 
EDSR.
In this model, a single electron occupies a one-dimensional (1D) parabolic QD, 
subject to a homogeneous static magnetic field $\mathbf B$, 
an inhomogeneous, disordered transverse magnetic field 
$\mathbf B_\perp(z) \perp \mathbf B$,
and an ac driving electric field $\mathbf E(t)$
(see Fig. \ref{fig:setup}). 
We focus on the regime of \emph{strong driving}, when the 
 amplitude $A$ of the electron's spatial oscillations, induced by the ac electric
 field, is comparable to the length $L$ of
the QD. 
We show that EDSR can be induced
in this system by subharmonic driving, i.e., if the excitation
frequency $\omega$ is an integer fraction (1/2, 1/3, etc) of the Larmor
frequency $\omega_L$.
At strong driving we find that
(i) the Rabi frequencies at the subharmonic resonances 
are comparable to the Rabi frequency of the fundamental resonance, and
(ii) at each subharmonic resonance, the Rabi frequency can be maximized
by setting the drive strength to an optimal finite value. 
Finally, our model is used to describe electrical control of 
a spin-valley qubit in a disordered carbon nanotube (CNT).

The motivation for this study is threefold. 
(1) Speed of manipulation is a central quantity in practical quantum
information processing. 
It is tempting to believe that in EDSR, a stronger electric drive
implies faster spin manipulation; in fact, we are unaware of any 
theoretical or experimental results indicating deviations from this relation. 
Our present study of the strong-driving regime allows us to reconsider (and
refute) this expectation, and to estimate the drive strength allowing
for the fastest spin control achievable in the strong-driving regime.
(2) Nonlinear processes arising from the interaction of matter and
electromagnetic fields are fundamentally important and have found
a wide range of applications in nonlinear optics as well as in nanoelectronics. 
Subharmonic EDSR is a distinct example of such nonlinear processes,
and might gain importance as a mechanism of nonlinear interaction between
single-spin QDs and microwave nanocircuits\cite{Trif,Petersson}.
(3) 
In a recent experiment using a CNT QD\cite{Laird}, 
the strong-driving regime of EDSR has been achieved:
the authors estimated\cite{Note1}
a maximum electric-field
amplitude of $E_{\rm ac} \approx 4 \times 10^4\ {\rm V}/{\rm m}$ ,
QD length $L \approx 100 $ nm, and
level spacing $\hbar \omega_0 \approx 3$ meV, 
implying a ratio $A/L \approx 1.3$ at
maximum drive power.
The same group has demonstrated\cite{FeiPei} the existence of
subharmonic EDSR, up to five-photon transitions, in CNT devices.
These experimental findings motivate the study of 
the strong-driving regime of EDSR.

\emph{Model.}
The static Hamiltonian, describing a single spinful electron in a 1D parabolic
QD, is defined as (see Fig. \ref{fig:setup}).
\bnen \label{H}
H_{stat} = \frac{p_z^2}{2m} + \frac 1 2 m \omega_0^2 z^2 + 
\frac 1 2 B \sigma_z + \frac 1 2 B_\perp(z)  \sigma_x,
\eden
where $m$ is the effective mass of the electron, $\omega_0$ is
the angular frequency associated to the harmonic confinement, 
$B \equiv \hbar \omega_L$ 
[$B_\perp(z)$] is the dc homogeneous longitudinal 
[disordered transverse] 
magnetic field 
pointing along the $z$ [$x$] direction,
and $\sigma_{x,z}$ are the first and third Pauli matrices representing
the electron spin (or pseudospin, see below).
The unit matrix in spin space is suppressed.
The QD confinement length is
$L=\sqrt{\hbar/m\omega_0}$.

In practice, the disordered transverse field (DTF) $B_\perp(z)$
might be induced by nuclear spins or other short-range impurities.
Therefore, we describe the DTF as a sum of Dirac deltas with
random prefactors:
\bnen
B_\perp (z) = a \sum_i \xi_i  \delta(z-z_i),
\eden
where $a$ is the lattice constant and $\xi_i$ are 
independent, identically distributed, and zero-mean 
random variables: $\langle \xi_i \rangle = 0$ and 
$\langle \xi_i \xi_j \rangle = \xi^2 \delta_{i,j}$. 
The dimension of $B$, $B_\perp$, $\xi_i$, and $\xi$ is energy.

The QD spreads over many impurity sites.
Therefore, due to the central limit theorem, 
the matrix elements of the DTF between the
harmonic oscillator basis functions $|n\rangle$ ($n\in \mathbb{N}$)
are well approximated by zero-mean Gaussian random variables
with standard deviation of the order of $\xi \sqrt{\frac a L}$.
To characterize the corresponding energy scale, we introduce
$\tilde{B}_\perp \equiv\sqrt{\big\langle\langle0|B_\perp (z)|0\rangle^2\big\rangle_{\rm dis}}=\xi\sqrt{\frac{a}{L\sqrt{2\pi}}}$,
where $\langle . \rangle_{\rm dis}$ is disorder averaging for the
realizations of the $\xi_i$'s.

To control the two-level system, we use an oscillating electric field
along the nanowire holding the QD:
\bnen \label{HE}
H_E = |e| E_{\rm ac} z \cos(\omega t).
\eden
We characterize 
the length scale of this driving by 
the amplitude 
$A = |e| E_{\rm ac} L^2/\hbar \omega_0$
of the center-of-mass oscillations of the
electron induced by the electric field.
We consider the experimentally motivated energy 
scale hierarchy
\bnen
\label{eq:hierarchy}
\hbar \omega_0 \gg B \sim \hbar \omega 
\gg \tilde B_\perp.
\eden

\emph{Tool: the co-moving frame.} 
Our goal is to describe the coherent spin dynamics in the 
strong-driving regime, i.e., when the electric field is strong enough
to induce an oscillation amplitude
comparable to the confinement length, $A\sim L$.
This condition corresponds to the 
energy-scale relation $|e| E_{\rm ac} L \sim \hbar \omega_0$.
In this regime, the electric field
cannot be treated as a small perturbation compared to the harmonic 
oscillator level spacing.
Instead of using perturbation theory, we transform the
Hamiltonian $H = H_{stat} + H_E$ into the wave-function basis 
co-moving with the confinement potential\cite{Rashba2008,SanJose-prb}.
This transformation is represented by the time-dependent unitary
operator $U(t)=\sum_n |n \rangle\langle n(t)|$,
where $\ket{n(t)}$ is the $n$-th harmonic oscillator eigenfunction 
of the instantaneous orbital Hamiltonian
$p_z^2/2m + m \omega_0^2 z^2/2 + H_E(t)$.
This transformation approximately 
decouples the ground-state spin doublet from the
excited states, thereby allowing us to analyze the spin dynamics 
using a $2\times 2$ effective Hamiltonian. 

Transformation to the co-moving frame yields the
Hamiltonian matrix 
$H'_{nm} = \bra{n} \left(UHU^\dag+i\hbar \dot U U^\dag\right) \ket{m}$,
which we express as 
\bnen
H'_{nm} =  
E_n (t) \delta_{nm}+\frac{1}{2}B\delta_{nm}\sigma_z+\frac{1}{2}\bar B^{nm}_\perp(t)\sigma_x-\varepsilon_{nm}(t).
\label{eq:comovingH}
\eden
Here, $E_n(t)$ is an electric-field-dependent shift of the orbital
energies.
The DTF, represented by 
$\bar B_\perp^{nm}(t) \equiv \bra{n(t)} B_\perp(z) \ket{m(t)}$,
 becomes time dependent after the transformation.
The spin-independent last term of Eq. \eqref{eq:comovingH} reads as
  \bnen
  \varepsilon_{nm}(t)=\frac{i \hbar\omega A}{\sqrt{2}L}\sin{\omega t}\left(\sqrt{n}\delta_{n,m+1}-\sqrt{m}\delta_{m,n+1}\right).
  \eden
  
In the co-moving frame, we can safely truncate $H'$ to
the ground-state spin subspace and use the effective spin Hamiltonian
$H'_{00}$ to describe spin dynamics. 
It is possible to take into account the coupling of this two-dimensional
subspace to higher-lying states via perturbation theory, yielding
spin-dependent corrections to $H'_{00}$ of the order of 
$\bar B_\perp^{01} \varepsilon_{10}/\hbar \omega_0 \sim 
\tilde B_\perp  (\omega/\omega_0) (A / L)$.
This implies that it is indeed justified to use $H'_{00}$ as the 
leading-order spin Hamiltonian as long as $A \ll L \omega_0 /\omega$,
which includes the case $A \sim L$ of our interest.

\emph{Rabi frequencies.}
We have concluded that the effective spin Hamiltonian is
\bnen
\label{eq:h00}
H'_{00} = \frac 1 2 B \sigma_z + \frac 1 2 \bar B_\perp(t) \sigma_x,
\eden
where 
$\bar B_\perp(t)\equiv \bra{0(t)} B_\perp(z) \ket{0(t)}$.
To express the $N$-photon Rabi frequency, we write 
$\bar B_\perp(t)$ in Fourier series:
\bean
\label{eq:fourier}
\bar B_\perp(t)&=&
\hbar \Omega^{(0)} + 
\sum_{N=1}^\infty 2\hbar\Omega^{(N)} \cos{\left(N\omega t\right)},
\\
\label{eq:rabiN}
\hbar\Omega^{(N)}&=&\frac{1}{2\pi}\int_0^{2\pi} d(\omega t)\bar B_\perp(t)\cos{\left(N\omega t\right)}.
\eean
The Fourier series \eqref{eq:fourier} contains  cosine terms only, as
$H_E(t)$ and hence $\bar B_\perp(t)$ 
are even functions of $t$.

In the rotating wave approximation\cite{Shirley}, the $N$-photon 
Rabi frequency, characterizing the spin-flip rate at 
driving frequency $\omega = \omega_L/N$, 
is simply given by $|\Omega^{(N)}|$.
Recall that $\Omega^{(N)}$ is itself a random variable, as its definition 
is based on the random variables $\xi_i$ building up the DTF.
We characterize the typical value of the Rabi frequency
by the standard deviation of $\Omega^{(N)}$, that is,
$
\sigma\left(\Omega^{(N)}\right)
\equiv
\sqrt{\big\langle\left[\Omega^{(N)}\right]^2\big\rangle_{\rm dis}}
$.
It can be expressed as 
\begin{eqnarray} 
\nonumber
\sigma\left(\Omega^{(N)}\right)
&=&
\frac{\tilde B_\perp}{2 \pi \hbar}
\left[
\int_0^{2\pi} d\tau \int_0^{2\pi} d\tau'
e^{-\frac{A^2}{2L^2}(\cos\tau-\cos\tau')^2}
\right.
\\
&\times & \left.
\cos{(N\tau)}
\cos{(N\tau')}
\right]^{1/2}.
\label{main}
\end{eqnarray}
From now on, we refer to the typical Rabi frequency simply as
the `Rabi frequency'.

\begin{figure}
\includegraphics[width=0.4\textwidth]{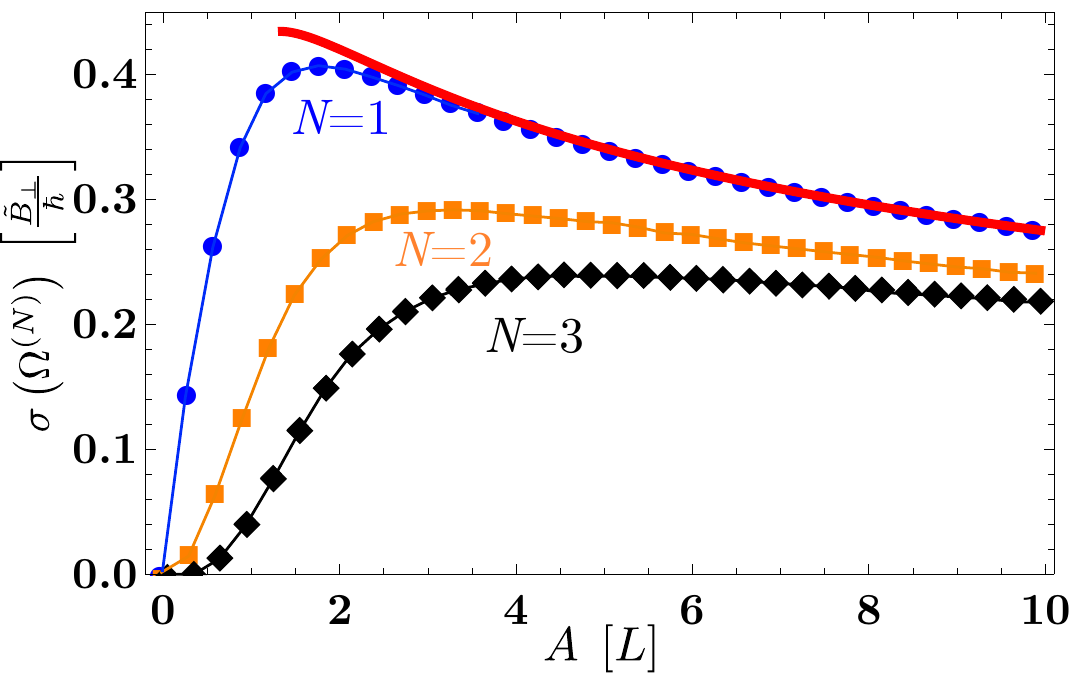}
\caption{\label{fig:Rabi}(Color online) The $N$-photon Rabi frequencies 
for $N=1,2,3$, as functions of the 
oscillation amplitude of the electronic wavefunction.  
Thick red line: the $A\gg L$ asymptote, Eq. \eqref{eq:AggL}.
}
\end{figure}

In Fig. \ref{fig:Rabi}, we plot the $N$-photon Rabi-frequencies, 
obtained by numerically integrating \eqref{main}, as a function of the  
oscillation amplitude $A$ of the electron wave packet. 
The main features in Fig. \ref{fig:Rabi} are as follows.
(i)  In the weak-driving regime $A\ll L$, the $N$-photon 
Rabi frequencies are proportional to $A^N \propto E_{\rm ac}^N$.
This is consistent with perturbation theory. 
The asymptote of \eqref{main} corresponding to this case is
$
\sigma\left(\Omega^{(N)}\right)\approx \frac{\tilde B_\perp}{\hbar}\frac{A^N}{L^N}\sqrt{\frac{(2N)!}{(2^NN!)^3}}$.
(ii) In the regime $A\gtrsim L$, the Rabi frequencies of the subharmonic
resonances $N=2,3$ are comparable to the Rabi frequency of the fundamental resonance 
$N=1$.
This is in sharp contrast to the behavior in the weak-driving regime.
(iii) Each Rabi-frequency curve has a maximum in the regime $A\sim L$, 
with the maximum points shifting to larger amplitudes $A$ as
the photon number $N$ is increased. 
(iv) After reaching their maxima, the Rabi frequencies decay
if the drive strength is increased to the $A > L$ regime. 
The asymptote of \eqref{main} for the case $N=1$ and $A \gg L$ 
reads as
\bnen
\label{eq:AggL}
\sigma\left(\Omega^{(1)}\right)
\approx \frac{ \tilde B_\perp}{\hbar} \sqrt{\frac{L}{A}}
\left[
\left(C^{(1)}_1+C^{(1)}_2\ln{\frac{A}{L}}\right)
\right]^{1/2},
\eden  
where $C^{(1)}_1=\frac{7\ln{2}-4+\gamma}{\sqrt{2\pi^3}}\approx0.18$ and $C^{(1)}_2=\sqrt{\frac{2}{\pi^3}}\approx0.25$, where $\gamma\approx0.58$ is the Euler-Mascheroni constant. The asymptotic result \eqref{eq:AggL} 
is shown as the red line in Fig. \ref{fig:Rabi}.

A simple interpretation of the decaying trend (iv) of the Rabi frequencies
for $A \gg L$ can be given using the simplified example of a zigzag-like
driving electric field.
To this end, we replace the harmonic driving $H_E \propto \cos \omega t$
by $H_E = |e| E_{\rm ac} z f(t)$ with
$f(t)$ being a piecewise linear function of
time, decreasing from 1 to -1 (increasing from -1 to 1) in the first (second) half  
of the period.
For this $f(t)$, the center of the 
electronic wave function moves with constant velocity
between turning points, hence it sweeps through any 
$L$-long segment of its orbit in time $\Delta t = \pi L / 2A\omega $.
First, this implies that Eq. \eqref{eq:rabiN} can be approximated by
(for the case $N=1$) 
\bean
\label{eq:approx}
\hbar \Omega^{(1)} \approx 
\frac 1 2 \frac L A \sum_{j=0}^{{\rm floor}(2A/L)} 
\bar B_\perp(j\Delta t)
\cos (\omega j \Delta t)
.
\eean
Second, it implies that 
it is reasonable to approximate the correlation time of 
$\bar{B}_\perp(t)$ with $\Delta t$; i.e., 
for $j\neq j'$, the distance between the center of the electronic wave function
at time $j\Delta t$ and $j'\Delta t$ is at least $L$, hence 
the corresponding values of $\bar B_\perp$ can be regarded as
uncorrelated: $\langle \bar B_\perp(j\Delta t) \bar B_\perp(j' \Delta t) \rangle_{\rm dis} \propto \delta_{j,j'}$.
Since the terms of the sum in Eq. \eqref{eq:approx} are uncorrelated,  
the standard deviation of the sum
varies with $A$ as $\sqrt{A/L}$, hence the standard deviation of the 
Rabi frequency
obeys $\sigma\left(\Omega^{(1)}\right) \propto \sqrt{L/A}$. 
Although the logarithmic correction obtained in Eq. \eqref{eq:AggL} is absent
in the case of this slightly modified driving profile $f(t)$, 
the decaying trend of the  
Rabi frequency with growing amplitude $A$ is indeed reproduced. 

An important practical consequence of the results shown
in Fig. \ref{fig:Rabi} is the following. 
If the fundamental or 
a subharmonic resonance is exploited with the aim of fast qubit control, then
it is not desirable to increase the amplitude $E_{\rm ac}$ 
of the driving electric field
as much as possible. Instead, there exists an optimal, finite value 
of $E_{\rm ac}$, of the order of $\hbar\omega_0/|e| L$, 
which maximizes the spin-flip rate for the given resonance.

\emph{Two-component DTF.}
The results obtained from our simple model \eqref{H}, 
where the DTF has only one Cartesian
component ($x$), can be easily generalized to the case where the DTF
has two components ($x$,$y$) transverse to the dc B-field. 
Then, the DTF Hamiltonian reads
$\frac{1}{2}\mathbf{B_\perp}(z)\boldsymbol\sigma$, where 
$\boldsymbol\sigma=(\sigma_x,\sigma_y,\sigma_z)$,
$
\mathbf{B}_\perp (z) = a \sum_i \boldsymbol\xi_i  \delta(z-z_i),
$
and $\boldsymbol \xi_i = (\xi_{i,x},\xi_{i,y},0)$.
Furthermore, 
$\langle \xi_{i,\alpha} \rangle_{\rm dis} = 0$ and  
$\langle \xi_{i,\alpha} \xi_{j,\beta} \rangle_{\rm dis}=\delta_{i,j} \delta_{\alpha,\beta}\xi^2/2$, with $\alpha,\beta \in \{x,y\}$. 
In this case of two-component DTF, 
the effective spin-driving field can be described by 
complex Fourier coefficients:
\begin{equation}
\Omega^{(N)}= \frac{1}{h} \int_0^{2\pi}d(\omega t)\langle0(t)|
B_{\perp,x}(z) + i B_{\perp,y} (z)
|0(t)\rangle\cos{(N\omega t)}.
\end{equation}
We identify the typical Rabi frequency with
$\sigma\left(\Omega^{(N)}\right)=\sqrt{\langle \left|\Omega^{(N)}\right|^2 \rangle_{\rm dis}}$.
With these new definitions, it is straightforward to show 
that the previously obtained results 
such as Eq. \eqref{main} and Fig. \ref{fig:Rabi} remain valid for
the case of a two-component DTF.

\emph{Application: coherent control of a spin-valley qubit in a disordered
CNT.} 
We show that the simple model developed above is applicable 
to describe the electrically driven dynamics of a spin-valley qubit
in a disordered CNT.
Electrically induced
qubit rotation in a similar system was observed\cite{Laird}, 
and it 
was attributed to the bent geometry\cite{FlensbergMarcus} 
of the sample; 
in contrast, following we describe an alternative mechanism 
that is active even in a straight CNT.

A simple model Hamiltonian of a single-electron parabolic QD in a
weakly disordered CNT in the presence of an external 
magnetic field $\boldsymbol{\mathcal B} = (\mathcal B_x,0,\mathcal B_z)$ reads as
\begin{eqnarray}
H_{\rm CNT}&=&\frac{p_z ^2}{2m} + \frac 1 2 m \omega_0^2 z^2 -
\frac{\Delta_{\rm so}}{2}s_z \tau_3+\frac{1}{2}\mathbf{b}(z)\boldsymbol\tau
\nonumber\\
&+&\frac{1}{2}g_{\rm s}\mu_B \left(\mathcal{B}_x s_x+\mathcal{B}_z s_z\right)+
\frac{1}{2}g_{\rm orb}\mu_B \mathcal{B}_z \tau_3.
\label{eq:cnt}
\end{eqnarray}  
The terms of $H_{\rm CNT}$ are, respectively:
electronic kinetic energy of the motion along the CNT axis ($z$);
parabolic confinement along $z$;
spin-orbit interaction characterized by the spin-orbit energy $\Delta_{\rm so}$;
valley-mixing short-range 
potential disorder with $\mathbf{b}(z) = (b_1(z),b_2(z),0)$;
spin Zeeman effect; 
and orbital Zeeman effect. 
Furthermore, $m$ is the electronic effective mass,
$\mathbf{s} = (s_x,s_y,s_z)$ [$\boldsymbol \tau = (\tau_1,\tau_2,\tau_3)$]
is the vector of Pauli matrices in the spin [valley]
space spanned by $\ket{ \! \uparrow}$ and $\ket{\! \downarrow}$ 
[$\ket{K}$ and $\ket{K'}$], 
and $g_{\rm s}$ ($g_{\rm orb}$) is the spin (orbital) g-factor. 
Identity matrices in spin and valley space are omitted.

The valley-mixing part of $H_{\rm CNT}$, 
describing short-range disorder reads
$\mathbf{b}(z)=\frac{\Omega_{\rm cell}}{4R\pi}\sum_i\boldsymbol\xi_i\delta(z-z_i)$,
where $\Omega_{\rm cell}$ is the unit cell area of the graphene 
lattice, $R$ is the
CNT radius, and $\boldsymbol \xi_i=(\xi_{i,1},\xi_{i,2},0)$ is a vector 
representing the impurity site $i$: $|\boldsymbol \xi_i|$ is the
random on-site energy on the impurity site, and the direction of 
$\boldsymbol \xi_i$ is set by the location of the impurity 
along the CNT circumference
\cite{Palyi-cnt-spinblockade,Palyi-valley-resonance}. 
Potential disorder appears in $H_{\rm CNT}$ only via the
valley-mixing term $\frac 1 2 \mathbf{b}(z) \boldsymbol \tau$; 
the valley-independent part is disregarded as it does not influence 
spin and valley dynamics. 

The basis states of the 
spin-valley qubit are defined as the ground-state doublet 
$\ket{K \! \uparrow},\ket{K' \! \downarrow}$
of  $H_{\rm CNT}$ in the absence of disorder ($\mathbf{b}(z) = 0$)
and external $B$-field ($\mathcal B_x = \mathcal B_z = 0$).
We describe the dynamics of the spin-valley qubit
induced by the simultaneous presence of disorder,
external $B$-field and electric driving, 
the latter being described by
$H_E$ as defined in Eq. \eqref{HE}.
We consider the case 
when $\hbar \omega_0 \gg \Delta_{\rm so}$ 
and $\Delta_{\rm so}$ exceeds the energy scales of
disorder ($\sqrt{\langle |\bra{0}  b_{1/2}(z) \ket{0} |^2 \rangle_{\rm dis}}$) and Zeeman splittings in $H_{\rm CNT}$. 

We derive an effective $2\times 2$ Hamiltonian 
for the spin-valley qubit, that is formally identical to 
 Eq. \eqref{eq:h00}.
 The derivation proceeds as follows. 
First, we transform the total Hamiltonian $H_{\rm CNT}+H_E$ to the
co-moving frame of the oscillating electron, as 
done earlier to obtain Eq. \eqref{eq:comovingH}.
Then, we truncate the Hilbert-space to the four-dimensional
subspace corresponding to the ground-state orbital level
of the parabolic confinement. 
Finally, we obtain a $2 \times 2$ Hamiltonian for the spin-valley qubit 
by decoupling its subspace from the subspace of the higher-lying doublet 
via second-order Schrieffer-Wolff perturbation theory 
\cite{Winkler,FlensbergMarcus,Szechenyi} 
in the disorder and Zeeman matrix elements. 
The resulting effective spin-valley qubit Hamiltonian,
expressed in the $\ket{K\uparrow}$, $\ket{K' \downarrow}$ basis, reads as
\bnen
H_{\rm sv} = \frac{1}{2}  (g_{\rm s} + g_{\rm orb}) \mu_B \mathcal B_z \sigma_z
+
\frac 1 2 \frac{g_{\rm s} \mu_B \mathcal  B_x}{\Delta_{\rm so}} \bra{0(t)} \mathbf{b} \ket{0(t)}
\boldsymbol \sigma.
\eden
The correspondence between this Hamiltonian and that of Eq. \eqref{eq:h00}
implies that the spin-valley qubit undergoes coherent Rabi oscillations
whenever
the $N$-photon resonance condition 
$ \hbar \omega N = \mu_B (g_{\rm s} + g_{\rm orb}) \mathcal B_z$ is fulfilled.
The corresponding Rabi frequency at the $N$-photon 
resonance is given by Eq. \eqref{main} and Fig. \ref{fig:Rabi}, 
with the substitution
\bnen
\label{eq:BperpCNT}
\tilde B_\perp\mapsto
\frac{g_{\rm s} \mu_B \mathcal B_x}{\Delta_{\rm so}} 
\sqrt{\frac{\Omega_{\rm cell}}{2(2\pi)^{3/2}RL}}\xi
\eden
and $\xi = \sqrt{\avg{\boldsymbol \xi_i^2}_{\rm dis}}$.
Note that Eq. \eqref{main} and Fig. \ref{fig:Rabi} apply in this case only if
the counterpart of the condition \eqref{eq:hierarchy} is fulfilled,
i.e., if $\mu_B(g_{\rm s}+g_{\rm orb}) B_z$ exceeds
$\tilde B_\perp$, the latter being defined by Eq. \eqref{eq:BperpCNT}.

For the realistic parameter set $R= 1$ nm, $L=100$ nm, 
$\xi = 22$ meV
(e.g., 50 impurity sites within the QD, each with a random $\pm 0.5$  eV on-site energy),
$\mathcal B_x=50$ mT,
$\mathcal B_z = 10$ mT,
$g_{\rm s} = 2$, $g_{\rm orb}=50$, and
$\Delta_{\rm so} = 0.5$ meV, 
the single-photon resonance frequency is $\omega \approx 46$ GHz,
and the maximal Rabi frequency at the single-photon 
resonance is $\sigma\left(\Omega^{(1)}\right) \approx 0.63$ GHz.

Finally, we point out that our 1D and 2D DTF models
are also applicable to describe
strongly driven EDSR of heavy holes in semiconductors
in the presence of Ising-type hyperfine interaction\cite{Fisher},
and 
electrically driven valley resonance in CNTs\cite{Palyi-valley-resonance},
respectively.

\emph{Note added:} A recent experiment\cite{Stehlik-edsr} 
revealed strong subharmonic
resonances in EDSR, similar to those predicted in this work, but 
presumably caused by a different mechanism (Landau-Zener transitions).


\begin{acknowledgments}
We thank G. Burkard and J.Romh\'anyi 
for useful discussions,
and especially E. Laird for questions inspiring this work. 
We acknowledge funding from 
the EU Marie Curie Career Integration Grant No. CIG-293834 (CarbonQubits),
the OTKA Grant No. PD 100373, and the EU GEOMDISS project.
A.~ P.~ is supported by the 
J\'anos Bolyai Scholarship of the Hungarian Academy of Sciences.
\end{acknowledgments}



\bibliography{edsr-manuscript}

\end{document}